\documentclass[12pt,twoside,a4paper,amsmath,amssymb,aps,showkeys,showpacs]{revtex4}

\usepackage{amsfonts}
\usepackage[russian]{babel}
\usepackage{graphics} 
\usepackage{epsfig}

\begin{document}
\DeclareGraphicsExtensions{.jpg,.pdf,.mps,.png,} 
\thispagestyle{myheadings}
\title{Generalized Dynamics of the Mass Point with Internal Degrees
of Freedom}

\author{A.N.Tarakanov}
\email{tarak-ph@mail.ru} \affiliation{Minsk State High
Radiotechnical College}

\begin{abstract}
An equation of motion of the mass point with internal degrees of
freedom in scalar potential $U$ depending on relative coordinates
and time, velocity and accelerations is obtained both for
non-relativistic and relativistic case. In non-relativistic case a
generalization of the energy conservation law follows, if
$\partial U / \partial t = 0$ fulfilled. A concept of work is
generalized to relativistic case, leading to corresponding
integral of motion, if $\partial U / \partial \tau = 0$ fulfilled,
where $\tau$ is proper time of the point. In neglecting an
internal degrees of freedom and absence of interaction this
integral of motion gives standard Special Relativity.
\end{abstract}


\maketitle
\section{Equation of Motion and the Energy Conservation}

A long period of supremacy of quantum theories did not crush an
interest in classical description of quantum systems. In this
connect some conclusions, following from the basic equation of
dynamics, the Second Newton's Law, should be noted. As it is well
known from the Helmholtz epoch (~\cite{Hel1}), the Second Newton's
Law for conservative systems

$$
 \frac{d{\bf P}}{dt} = {\bf F} \eqno{({\rm I}.1)}
$$
\noindent gives a force acting at the mass point in the form ${\bf
F} = - \nabla U = - \partial U / \partial {\bf R}$, where $U =
U({\bf R})$ is potential function of coordinate of the mass point.
As a result Eq.(I.1) and a definition of elementary work

$$
dA = ({\bf F} \cdot d{\bf R}) = (\frac{d{\bf P}}{dt} \cdot d{\bf
R}) = ({\bf V} \cdot d{\bf P}) \;, \eqno{({\rm I}.2)}
$$
\noindent where \textbf{R} and ${\bf V} = d{\bf R} / dt$ are
respectively radius vector and velocity of the mass point relative
to origin of coordinate system, coupled with absolute rest
reference frame (r.f.), give a conservation of total mechanical
energy

$$
E = \frac{m{\rm {\bf V}}^2}{2} + U({\rm {\bf R}}) \;. \eqno{({\rm
I}.3)}
$$
\noindent As soon as absolute r.f. be coupled with any physical
object, one can say that the motion of mass point takes place in
the field, created by this object and characterized by potential
function $U({\bf R})$.

It is clear from the common considerations that a motion of mass
point in the field of some object should determined by potential
function depending not only on relative coordinates \textbf{R},
but also at least on relative velocity \textbf{V} and
accelerations, as well as on time, so that $U = U(t,{\bf R},{\bf
V},{\bf W},{\bf \dot {W}},...,{\bf W}^{(N)})$, where ${\bf
W}^{(k)} = d^{k}{\bf W} / dt^{k}$, a time dependence being
specified by internal dynamics of the mentioned object. Remember
W.Weber (~\cite{Web1}-~\cite{Web2}), who tried to explain
electrical phenomena as a result of electric interaction of
elementary particles, so called \textit{electric atoms}, depending
both on their relative disposition \textbf{R} and on their
relative velocity \textbf{V} and acceleration ${\bf W} = d{\bf V}
/ dt$.

In this case corollaries ${\bf F} = - \nabla U$ and (I.3) from
equation of motion (I.1) should be changed forasmuch as total
differential of the function $U$ is

$$
dU = \frac{\partial U}{\partial t}dt + (\frac{\partial U}{\partial
{\bf R}} \cdot d{\bf R}) + (\frac{\partial U}{\partial {\bf V}}
\cdot d{\bf V}) + \sum\limits_{k = 0}^N {(\frac{\partial
U}{\partial {\bf W}^{(k)}} \cdot d{\bf W}^{(k)})} \;. \eqno{({\rm
I}.4)}
$$

Indeed, definition of elementary work of the force (I.2) gives
more general expression for force, namely

$$
{\bf F} = - \frac{\partial U}{\partial {\bf R}} + [{\bf C}\times
{\bf V}] \;, \eqno{({\rm I}.5)}
$$
\noindent where \textbf{C} is some pseudo-vector, inherent in mass
point, and additional term $[{\bf C}\times {\bf V}]$ having a
sense of gyroscopic force. As far back as Helmholtz in his work
"On the conservation of force" (~\cite{Hel1}; Addition 3) pointed
out at formula (I.6).

Furthermore, when interaction takes place the momentum vector
\textbf{P} has a meaning of dynamical momentum. It can be written
as a sum of kinematical momentum $m{\bf V}$ and some addition
\textbf{A} (a vector potential), connected both with internal
structure of mass point, and with interaction

$$
{\bf P} = m{\bf V} + {\bf A}\;. \eqno{({\rm I}.6)}
$$

Then Eqs.(I.2) and (I.6) give

$$
dA = ({\bf F} \cdot d{\bf R}) = ({\bf V} \cdot d(m{\bf V} + {\bf
A})) = d\left( {\frac{m{\bf V}^2}{2} + ({\bf A} \cdot {\bf V})}
\right) - ({\bf A} \cdot d{\bf V}) \;, \eqno{({\rm I}.7)}
$$
\noindent or

$$
 d\left( {\frac{m{\bf V}^2}{2} + ({\bf A} \cdot {\bf V})}
\right) + (\frac{\partial U}{\partial {\bf R}} \cdot d{\bf R}) -
({\bf A} \cdot d{\bf V}) =
$$
$$
 = d\left( {\frac{m{\bf V}^2}{2} + ({\bf A} \cdot {\bf V})
+ U(t,{\bf R},{\bf V},{\bf W},{\bf \dot {W}},...,{\bf W}^{(N)})}
\right) -
$$
$$
 - \frac{\partial U}{\partial t}dt - (\frac{\partial U}{\partial {\bf
V}} \cdot d{\bf V}) - ({\bf A} \cdot d{\bf V}) - \sum\limits_{k =
0}^N {(\frac{\partial U}{\partial {\bf W}^{(k)}} \cdot d{\bf
W}^{(k)})} = 0\,\,. \eqno{({\rm I}.8)}
$$

Now, if one suppose

$$
{\bf A} = - \frac{\partial U}{\partial {\bf V}} + [{\bf S}\times
{\bf W}] \;, \eqno{({\rm I}.9)}
$$
\noindent where \textbf{S} is some pseudo-vector, coupled with the
mass point considered, \textbf{W} is an acceleration of this
point, then dynamical momentum (I.6) will get an expression

$$
{\bf P} = m{\bf V} - \frac{\partial U}{\partial {\bf V}} + [{\bf
S}\times {\bf W}] \;, \eqno{({\rm I}.10)}
$$
\noindent whereas it follows from Eq.(I.8)

$$
 \frac{dE}{dt} = \frac{\partial U}{\partial t} + \sum\limits_{k =
 0}^N {(\frac{\partial U}{\partial {\bf W}^{(k)}} \cdot {\bf
 W}^{(k + 1)})} \;,\eqno{({\rm I}.11)}
$$
\noindent where quantity

$$
E = \frac{m{\bf V}^2}{2} + ({\bf V} \cdot [{\bf S}\times {\bf W}])
- ({\bf V} \cdot \frac{\partial U}{\partial {\bf V}}) + U(t,{\bf
R},{\bf V},{\bf W},{\bf \dot {W}},...,{\bf W}^{(N)}) \eqno{({\rm
I}.12)}
$$
\noindent is a generalization of Eq.(I.3) for total mechanical
energy. So, apart from standard kinetic and potential energies an
additional energy arises due to both internal degrees of freedom
and a dependence of potential energy on relative velocity.

Provided the condition

$$
\frac{\partial U}{\partial t} + \sum\limits_{k = 0}^N
{(\frac{\partial U}{\partial {\bf W}^{(k)}} \cdot {\bf W}^{(k +
1)})} = 0 \eqno{({\rm I}.13)}
$$
\noindent is fulfilled, the energy (I.12) will be an integral of
motion. Condition $dE / dt > 0$ corresponds to absorption of
energy by a mass point, and $dE / dt < 0$ corresponds to radiation
of energy.

In view of stated above, the equation of motion (I.1) should be
written down in the form

$$
\frac{d}{dt}\left( {m{\bf V} + [{\bf S}\times {\bf W}]} \right) -
[{\bf C}\times {\bf V}] = \frac{d}{dt}\frac{\partial U}{\partial
{\bf V}} - \frac{\partial U}{\partial {\bf R}} \;. \eqno{({\rm
I}.14)}
$$

Let's note here that derivatives of potential function with
respect to accelerations ${\bf W}^{(k)}$ do not enter into an
equation of motion. Therefore one can be restricted to dependence
of potential function only on acceleration \textbf{W}: $U =
U(t,{\bf R},{\bf V},{\bf W})$.

Equation (I.14) gives a number of special cases.

1. First Newton's law ($U(t,{\bf R},{\bf V},{\bf W}) = 0$, ${\bf
V} = {\bf const})$ takes place, if the relation

$$
\frac{d}{dt}[{\bf S}\times {\bf W}] = [{\bf C}\times {\bf V}] \;,
\eqno{({\rm I}.15)}
$$
\noindent including also trivial absence of internal structure,
${\bf C} = {\bf 0}$ \;, ${\bf S} = {\bf 0}$ \;, is fulfilled.

2. If the force (I.5), acting at a mass point, becomes zero, i.e.
the relation

$$
\frac{\partial U}{\partial {\bf R}} = [{\bf C}\times {\bf V}]
\eqno{({\rm I}.16)}
$$
\noindent is fulfilled, then dynamical momentum (I.10) will be a
conserved vector.

If besides the relation

$$
\frac{\partial U}{\partial {\bf V}} = [{\bf S}\times {\bf W}] \;,
\eqno{({\rm I}.17)}
$$
\noindent is fulfilled, then dynamical momentum coincides with
kinematical momentum and there take place uniform and rectilinear
motion.

For free mass point, when interaction is neglected, $U(t,{\bf
R},{\bf V},{\bf W}) = 0$, conservation law of dynamical momentum
takes place only if $[{\bf C}\times {\bf V}] = {\bf 0}$, and
uniform and rectilinear motion is a result of additional condition
$[{\bf S}\times {\bf W}] = {\bf 0}$.

3. If $\partial U / \partial {\bf R} = {\bf 0}$ and the relation

$$
[{\bf C}\times \left( {[{\bf S}\times {\bf W}] - \frac{\partial
U}{\partial {\bf V}}} \right)] = 0 \eqno{({\rm I}.18)}
$$
\noindent takes place, equation of motion (I.14) takes the form

$$
\frac{d}{dt}\left( {m{\bf V} + [{\bf S}\times {\bf W}] -
\frac{\partial U}{\partial {\bf V}}} \right) - \frac{1}{m}[{\bf
C}\times \left( {m{\bf V} + [{\bf S}\times {\bf W}] -
\frac{\partial U}{\partial {\bf V}}} \right)] = 0 \;, \eqno{({\rm
I}.19)}
$$
\noindent from which it follows, that dynamical momentum (I.10) is
precessing round a direction of pseudovector \textbf{C} with
angular velocity ${\bf \omega} = {\bf C} / m$.

In general case the relation (I.15) or relations (I.16) and (I.17)
necessary for performance of the first Newton's law, may not be
satisfied. Therefore for mass points with internal degrees of
freedom the inertia law in the form in which it has been
formulated by Galiley and Newton, has no place and cannot be
accepted as the first principle underlying mechanics. One can
generalize it in the following way: \textit{Mass point (body) with
internal degrees of freedom, given to itself, moves according to
equation (I.15) in which} $U(t,{\bf R},{\bf V},{\bf W}) = 0$.

\section{The equation of moments for a mass point with internal degrees of
freedom}

The equation (I.14) is insufficient for description of dynamics of
physical system. There is necessary in addition an equation of
moments, which for structureless mass point looks like $d{\bf L} /
dt = {\bf M}$, where ${\bf L} = [{\bf R}\times {\bf P}] = m[{\bf
R}\times {\bf V}]$ is angular momentum, ${\bf M} = [{\bf R}\times
{\bf F}]$ is total moment of external forces, acting at the
system. For individual mass point equation of moments follows from
the Eq.(I.1).

For a mass point with internal degrees of freedom, describing by
Eq.(I.1), in which force and momentum are specified by equations
(I.5) and (I.10), respectively, we have the relation

$$
 [{\bf R}\times \frac{d{\bf P}}{dt}] = \frac{d}{dt}[{\bf
R}\times {\bf P}] - [{\bf V}\times ( - \frac{\partial U}{\partial
{\bf V}} + [{\bf S}\times {\bf W}])] = - [{\bf R}\times
\frac{\partial U}{\partial {\bf R}}] + [{\bf R}\times [{\bf
C}\times {\bf V}]]\,\,, \eqno{({\rm II}.1)}
$$
\noindent implying the following equation of moments

$$
\frac{d{\bf L}}{dt} = {\bf M} + {\bf T}, \eqno{({\rm II}.2)}
$$
\noindent where

$$
{\bf L} \doteq [{\bf R}\times {\bf P}] = m[{\bf R}\times {\bf V}]
- [{\bf R}\times \frac{\partial U}{\partial {\bf V}}] + [{\bf
R}\times [{\bf S}\times {\bf W}]] \eqno{({\rm II}.3)}
$$
\noindent is a dynamical angular momentum,

$$
{\bf M} \doteq [{\bf R}\times {\bf F}] = - [{\bf R}\times
\frac{\partial U}{\partial {\bf R}}] + [{\bf R}\times [{\bf
C}\times {\bf V}]] \eqno{({\rm II}.4)}
$$
\noindent is a moment of force, acting at the mass point,

$$
{\bf T} \doteq [{\bf V}\times {\bf P}] = - [{\bf V}\times
\frac{\partial U}{\partial {\bf V}}] + [{\bf V}\times [{\bf
S}\times {\bf W}]] \eqno{({\rm II}.5)}
$$

\noindent is an additional twisting moment, or torque. In standard
mechanics the concept "torque" \, is applied to the moment of
force (II.4). Here we distinguish the moment of force (II.4) and
torque (II.5).

It should be noticed, that in the same way both equation $d{\bf L}
/ dt = {\bf M}$ follows from Eq.(I.1) for usual mass point and
equation (II.2) follows from Eq.(I.14) (i.e. Eq.(I.1), in which
\textbf{F} and \textbf{P} are specified by equations (I.5) and
(I.10)) for a mass point with internal degrees of freedom.

Solution of equation (I.14) may be obtained in principle, if
potential function $U = U(t,{\bf R},{\bf V},{\bf W})$ and time
dependence of pseudo-vectors \textbf{S} è \textbf{C}, coupled with
internal structure of mass point, are known. As it is known, one
of internal property of particles is spin, associated classically
with proper angular momentum of particle. Therefore a temptation
arises to connect pseudo-vectors \textbf{S} and \textbf{C} with
spin. However, having only definition (II.3) for angular momentum
it is impossible to define a concept of proper angular momentum.
Therefore pseudo-vectors \textbf{S}, \textbf{C} and their
equations of motion should either postulated here artificially or
determined starting from additional arguments. In particular, one
may go by the same way as a solid body in mechanics considered as
a system of mass point. Then it is possible to define a concept of
\textit{particle} with internal degrees of freedom as a system of
the same mass points, whose proper angular momentum is determined
relative to center of inertia of particle. Such procedure will be
made elsewhere. Here it is reasonably to generalize equations and
concepts above to relativistic case.

\section{Relativistic equation of motion}

Relativistic generalization of the second Newton's law for mass point is

$$
\frac{d\mbox{P}}{d\lambda} = \frac{1}{c}\mbox{F} \eqno{({\rm
III}.1)}
$$

\noindent where $\mbox{P} = \{P^{\mu} \} = (P^0,{\bf P})$,
$\mbox{F} = \{F^{\mu} \} = (F^0,{\bf F})$, $\mu = 0,1,2,3$, are
relativistic generalizations of momentum and force,
\textit{$\lambda$} is invariant parameter determined by the
interval

$$
dS^2 = \eta _{\mu\nu} dR^\mu dR^\nu = (dR^0)^2 - d{\bf R}^2 =
\sigma d\lambda ^2 \,, \quad \sigma = \pm 1 \;, \eqno{({\rm
III}.2)}
$$

\noindent where $\eta _{\mu\nu} = diag(1,-1,-1,-1)$. Thus, for
$\sigma = +1$ parameter $\lambda / c = \tau $ is a proper time of
"concomitant observer" \, $\mathrm{K'}$ moving together with event
defined by four-dimensional radius-vector $\mbox{R} = \{R^\mu \} =
(R^0,{\bf R})$. For $\sigma = - 1$ parameter $\lambda = S$
coincides with length of arc of world line of the event R.

Let's note an important fact, that standard Special Relativity
with interval (III.2) is valid exceptionally for inertial r.f.
Usually interval (III.2) is considered as a definition of distance
between two points in the Minkowski space ${\bf E}_{1,3}^{\rm R}$.
Then coordinates of a point in ${\bf E}_{1,3}^{\rm R}$, defined by
radius-vector R, are quantities relative to origin, coinciding
with origin of the rest inertial r.f., and have absolute character
in the meaning of absolute time and absolute space of Newton's
mechanics. Relative character in the meaning of Special Relativity
they acquire when interval (III.2) is coupled with r.f.
$\mathrm{K'}$, moving relative to the rest r.f. K with velocity
${\bf V} = cd{\bf R} / dR^0$. In this case radius-vector R is said
to be an event R, whose world line is a trajectory of the origin
of inertial r.f. $\mathrm{K'}$, moving with velocity \textbf{V} in
${\bf E}_{1,3}^{\rm R}$, i.e. in the space of the rest r.f. K.

For inertially moving r.f. $\mathrm{K'}$ r.h.s. of Eq.(III.1)
vanishes, and we obtain conservation of 4-momentum, whence it
follows conservation of

$$
\mbox{P}^2 = \eta _{\mu\nu} P^\mu P^\nu = (P^0)^2 - {\bf P}^2 =
\sigma m_0^2 c^2 \; , \eqno{({\rm III}.3)}
$$
\noindent if 4-momentum is defined as

$$
P^\mu = m_0 cU^\mu = m_0 cdR^\mu / d\lambda = m_0 dR^\mu / d\tau
\;. \eqno{({\rm III}.4)}
$$

Relations (III.3)-(III.4) are standard relations of Special
Relativity for kinematical momentum, which are extended on any
asymptotically free physical systems without any reason. Between
other things one may consider an expression (III.2) for
relativistic interval as a corollary from relations (III.3),
postulating connection between energy and momentum.

If some force be acting on moving r.f. $\mathrm{K'}$, the latter
is no longer inertial one. Then 4-momentum in Eq.(III.1) becomes
dynamical momentum, whose definition ought to be analogous to
Eq.(I.6)

$$
P^\mu = m_0 c\frac{dR^\mu}{d\lambda} + K^\mu \;, \eqno{({\rm
III}.5)}
$$

\noindent where $K^\mu$ is some addition to kinematical 4-momentum
(III.4) due to interaction between moving r.f. $\mathrm{K'}$ and
rest r.f. K.

In Newton's mechanics an interaction force (I.5) between
$\mathrm{K'}$ and K is determined by means of elementary work
(I.2) which may be written as $dA = - \eta _{ij} F^{i} dR^{j}$.
This work is a scalar under Galilei transformations, i.e. it is
the same in all non-relativistic inertial r.f., but it is not
covariant under Lorentz transformations.

Indeed, let $L_{\,.\,\nu }^{\mu}$ be matrix elements of the
Lorentz transformation $d{R'}^\mu = L_{\,.\,\nu }^{\mu} dR^{\nu}$,
satisfying to condition $\eta _{\lambda\kappa} L_{\,.\,\mu
}^{\lambda} L_{\,.\,\nu }^{\kappa} = \eta _{\mu\nu}$, so that
(~\cite{Tar})

$$
L_{\,.\,0}^0 = \gamma_\sigma = (1 - {\bf B}_0^{2\sigma})^{-1/2}
\;, \; L_{\,.\,i}^0 = \Gamma_{\sigma} V_{0i} / V_0 \;, \;
L_{\,.\,0}^i = - \Gamma_{\sigma} V_0^i / V_0 \;, \; L_{\,.\,j}^i =
\delta_{\,.\,j}^i - \frac{\gamma_{\sigma} - 1}{{\bf V}_0^2}
V_0^{i} V_{0j} \;, \eqno{({\rm III}.6)}
$$
\noindent where

$$
{\bf B}_0 = {\bf V}_0 / c \;, \quad V_0 = \vert {\bf V}_0 \vert =
\sqrt {{\bf V}_0^2} \;, \quad {\rm B}_0 = \vert {\bf B}_0 \vert =
\sqrt {{\bf B}_0^2} = c V_0 \;; \eqno{({\rm III}.7)}
$$
$$
\Gamma_\sigma = {\rm B}_0^{\sigma} \gamma_{\sigma} \;, \quad
\Gamma_{+} = {\rm B}_0 \gamma_{+} = {\rm B}_0 (1 - {\bf B}_0^2
)^{-1/2} \;, \quad \Gamma_{-} = \gamma_{-} / {\rm B}_0 = ({\bf
B}_0^2 - 1)^{-1/2} \;, \eqno{({\rm III}.8)}
$$
\noindent ${\bf V}_0$ is velocity of r.f. $\mathrm{K'}$ relative
to r.f. K.

Then the Lorentz transformation takes form

$$
dR'^0 = \gamma_{\sigma} \left[ dR^0 - \frac{{\rm B}_0^{\sigma}
({\bf V}_0 \cdot d{\bf R})}{V_0} \right] \;, \eqno{({\rm III}.9)}
$$
$$
d{\bf R'} = d{\bf R} + \left[ (\gamma_{\sigma} - 1) \frac{({\bf
V}_0 \cdot d{\bf R})}{c {\rm B}_0^{\sigma} {\rm B}_0} -
\gamma_{\sigma} dR^0 \right] \frac{{\rm B}_0^{\sigma}}{c {\rm
B}_0} {\bf V}_0 \;. \eqno{({\rm III}.10)}
$$

Transformation law of relativistic force looks as

$$
F'^\mu = L_{\,.\,\nu}^{\mu} F^{\nu} = L_{\,.\,0}^\mu F^0 +
L_{\,.\,i}^{\mu} F^i \;, \eqno{({\rm III}.11)}
$$
$$
F'^0 = \gamma_{\sigma} \left[ F^0 - \frac{{\rm B}_0^{\sigma} ({\bf
F} \cdot {\bf V}_0)}{V_0} \right] \;, \eqno{({\rm III}.12)}
$$
$$
{\bf F'} = {\bf F} + \left[ \frac{(\gamma_{\sigma} - 1)({\bf F}
\cdot {\bf V}_0 )}{{\bf V}_0^2} - \frac{\gamma_{\sigma} {\rm
B}_0^{\sigma} F^0}{V_0 } \right]{\bf V}_0 \;. \eqno{({\rm
III}.13)}
$$

Hence Eqs.(III.10) and (III.13) give transformation law for
elementary work

$$
 dA' = ({\bf F'} \cdot d{\bf R'}) = - \eta_{ij}
F'^{i} dR'^{j} = - \eta_{ij} L_{\,.\,\mu}^{i} L_{\,.\,\nu}^{j}
F^{\mu} dR^{\nu} =
$$
$$
 = dA + \Gamma_{\sigma}^2 \left[ 1 - \frac{({\bf V} \cdot {\bf
V}_0 )}{c^2{\rm B}_0 {\rm B}_0^{\sigma}} \right] c F^0 dt -
\frac{\Gamma_{\sigma}^2}{{\rm B}_0 {\rm B}_0^{\sigma}} ({\bf F}
\cdot {\bf V}_0 ) dt + \frac{\gamma_{\sigma}^2 - 1}{{\bf V}_0^2}
({\bf F} \cdot {\bf V}_0)({\bf V} \cdot {\bf V}_0) dt \;,
\eqno{({\rm III}.14)}
$$
\noindent whence it follows relativistic transformation of power
$N = cdA / dR^0 = ({\bf F} \cdot {\bf V})$

$$
 N' = c\frac{dA'}{dR'^0} = ({\bf F'} \cdot {\bf V'}) =
$$
$$
 = \frac{N + L_{\,.\,0}^0 L_{\,.\,i}^0 F^{0} V^{i} + c
 L_{\,.\,i}^0 L_{\,.\,0}^0 F^{i} + L_{\,.\,i}^0 L_{\,.\,j}^0
 F^{i} V^{j} + c [(L_{\,.\,0}^0)^2 - 1] F^0}{L_{\,.\,0}^0 +
 L_{\,.\,i}^0 V^{i} / c} =
$$
$$
 = \frac{N + c\Gamma_{\sigma}^2 \left[ 1 - \frac{({\bf V} \cdot
 {\bf V}_0)}{c^2 {\rm B}_0 {\rm B}_0^{\sigma}} \right] F^0 -
 \frac{\Gamma_{\sigma}^2}{{\rm B}_0 {\rm B}_0^{\sigma}}({\bf F}
 \cdot {\bf V}_0) + \frac{\gamma_{\sigma}^2 - 1}{{\bf V}_0^2}
 ({\bf F} \cdot {\bf V}_0)({\bf V} \cdot {\bf V}_0)}
 {\gamma_{\sigma} \left[ 1 - \frac{{\rm B}_0^{\sigma} ({\bf V}
 \cdot {\bf V}_0)}{V_0} \right]} \;. \eqno{({\rm III}.15)}
$$

Noncovariance of expression (I.2) and force transformation law
(III.13) are inconsistent with principle of relativity, whose
successive application means that equations and quantities, such
as scalars, 4-vectors, tensors etc., should be covariant under
Lorentz transformations in any theory. An expression, being a
scalar in some inertial r.f., ought to be scalar in another
inertial r.f. Therefore definition (I.2) should be generalized in
the form

$$
dW = - \eta_{\mu\nu} F^{\mu} dR^{\nu} = - F^0 dR^0 + dA = ( - cF^0
+ N)dt \;. \eqno{({\rm III}.16)}
$$

In standard Special Relativity, dealing with interval (III.2),
$F^0$ is defined from Eq.(III.1), where $P^0 = m_0 c dR^0 / d \tau
= m_0 c^2 \gamma = m_0 c^2 (1 - {\bf V}^2 / c^2)^{- 1/2}$,
\textbf{V} is absolute velocity of the mass point acquiring
acceleration ${\bf W} = d{\bf V} / dt$ under action of the force
\textbf{F}. Then taking into account the relation

$$
\frac{d\gamma}{dt} = \frac{\gamma^3}{c^2}({\bf V} \cdot {\bf W})
\;, \eqno{({\rm III}.17)}
$$
\noindent we obtain

$$
{\bf F} = \frac{d{\bf P}}{d\tau } = \gamma \frac{d(m_0 \gamma {\bf
V})}{dt} = m_0 \gamma^2 \left[ {\bf W} + \frac{\gamma^2}{c^2}({\bf
V} \cdot {\bf W}){\bf V} \right] \;, \eqno{({\rm III}.18)}
$$
$$
({\bf F} \cdot {\bf V}) = m_0 \gamma^4 ({\bf V} \cdot {\bf W}) \;,
\eqno{({\rm III}.19)}
$$
$$
F^0 = \frac{dP^0}{d\tau} = m_0 c \frac{d\gamma}{d\tau} = m_0 c
\gamma \frac{d\gamma}{dt} = \frac{m_0 \gamma^4}{c}({\bf V} \cdot
{\bf W}) = \frac{1}{c}({\bf F} \cdot {\bf V}) = \frac{N}{c} \;.
\eqno{({\rm III}.20)}
$$

Comparison of Eq.(III.16) with Eq.(III.20) shows that scalar
\textit{dW} is identically zero in all inertial r.f. However,
should moving r.f. $\mathrm{K'}$ be coupled with considered mass
point, the latter ceases to be inertial one. Then in such r.f. an
expression (III.2) for length of arc of the world line of the mass
point and relations (III.18)-(III.20) become invalid. It means
that scalar $dW =$ $= - \eta_{\mu\nu} F^{\mu} dR^{\nu} = -
\eta_{\mu\nu} F'^{\mu} dR'^{\nu}$ does not equal to zero,
conserving its covariant expression in all inertial r.f.
Relativistic force, acting at rest mass point in r.f.
$\mathrm{K'}$, may be expressed in terms of potential function $U
= U(\lambda, R^{\mu}, U^{\mu}, W^{\mu},
\dot{W}^{\mu},...,(W^{(N)})^{\mu})$ by analogy with (I.5) in
non-relativistic mechanics, where $R^\mu$, $U^\mu$, $W^\mu =
dU^\mu / d\lambda$, $(W^{(k)})^{\mu} = d^{k}W^\mu / d\lambda^k$
are relative radius-vector, 4-velocity and 4-accelerations of r.f.
$\mathrm{K'}$ relative to r.f. K. Forasmuch as relation
$\eta_{\mu\nu} U^{\mu} U^{\nu} = \sigma$ following from (III.2)
becomes invalid, arguments of potential function should considered
as independent variables, so that total differential of \textit{U}
equals

$$
dU = \frac{\partial U}{\partial \lambda} d\lambda + \frac{\partial
U}{\partial R^\mu} dR^{\mu} + \frac{\partial U}{\partial U^\mu}
dU^{\mu} + \sum\limits_{k = 0}^N \frac{\partial U}{\partial
(W^{(k)})^{\mu}} d(W^{(k)})^{\mu} \;. \eqno{({\rm III}.21)}
$$

Formulae (I.10) and (I.5) should considered as non-relativistic
limits of 4-momentum and 4-force

$$
P^{\mu} = m_0 c U^{\mu} - \eta^{\mu\nu} \frac{\partial U}{c
\partial U^{\nu}} + \frac{1}{2} \eta^{\mu\nu}
\varepsilon_{\nu\lambda\kappa\rho} S^{\lambda\kappa} W^{\rho} \;,
\eqno{({\rm III}.22)}
$$
$$
F^{\mu} = - \eta^{\mu\nu} \frac{\partial U}{\partial R^{\nu}} +
\frac{1}{2} \eta^{\mu\nu} \varepsilon_{\nu\lambda\kappa\rho}
C^{\lambda\kappa} U^{\rho} , \eqno{({\rm III}.23)}
$$
\noindent respectively, where $S_{\lambda\kappa}$ and
$C_{\lambda\kappa}$ are some antisymmetric tensors, characterizing
internal structure of the mass point.

Substitution of Eqs.(III.22)-(III.23) into equation (III.1) gives
next equation of motion

$$
\frac{d}{d\lambda} \left[ m_0 c U^{\mu} + \frac{1}{2}
\eta^{\mu\nu} \varepsilon_{\nu\lambda\kappa\rho} S^{\lambda\kappa}
W^{\rho} \right] - \frac{1}{2c} \eta^{\mu\nu}
\varepsilon_{\nu\lambda\kappa\rho} C^{\lambda\kappa} U^{\rho} =
\frac{1}{c} \eta^{\mu \nu } \left[ \frac{d}{d\lambda}
\frac{\partial U}{\partial U^{\nu}} - \frac{\partial U}{\partial
R^{\nu}} \right] \;. \eqno{({\rm III}.24)}
$$

Substitution of Eqs.(III.22)-(III.23) into equation (III.1) gives

$$
c \eta_{\mu\nu} \frac{dP^{\mu}}{d\lambda} dR^{\nu} = c
\eta_{\mu\nu} U^{\mu} dP^{\nu} = \eta_{\mu\nu} F^{\mu} dR^{\nu} =
\eta_{\mu\nu} U^{\mu} F^{\nu} d\lambda \;, \eqno{({\rm III}.25)}
$$
\noindent or

$$
\eta_{\mu\nu} U^{\mu} \frac{d}{d\lambda}\left[ m_0 c^2 U^\nu +
\frac{c}{2} \eta^{\mu\nu} \varepsilon_{\nu\lambda\kappa\rho}
S^{\lambda\kappa} W^{\rho} \right] = U^{\mu} \frac{d}{d\lambda}
\frac{\partial U}{\partial U^{\mu}} - U^{\mu} \frac{\partial
U}{\partial R^{\mu}} \;. \eqno{({\rm III}.26)}
$$

Hence we obtain equation

$$
\frac{dE}{d\lambda} = \frac{\partial U}{\partial \lambda} +
\sum\limits_{k = 0}^N \frac{\partial U}{\partial (W^{(k)})^{\mu}}
(W^{(k+1)})^{\mu} \;, \eqno{({\rm III}.27)}
$$
\noindent where quantity

$$
E = \frac{m_0 c^2}{2} \eta _{\mu\nu} U^{\mu} U^{\nu} + \frac{c}{2}
\varepsilon_{\mu\nu\lambda\kappa} U^{\mu} W^{\nu}
S^{\lambda\kappa} + U - U^{\mu} \frac{\partial U}{\partial
U^{\mu}} = \frac{m_0 c^2}{2} \sigma = const \;, \eqno{({\rm
III}.28)}
$$
\noindent is an integral of motion provided a condition

$$
\frac{\partial U}{\partial \lambda} + \sum\limits_{k = 0}^N
\frac{\partial U}{\partial (W^{(k)})^{\mu}} (W^{(k+1)})^{\mu} = 0
\eqno{({\rm III}.29)}
$$
\noindent is satisfied.

Neglecting internal structure of mass point and its interaction,
$U = 0$, from (III.28) we obtain $\eta_{\mu\nu} U^{\mu} U^{\nu} =
\sigma$, and expression (III.2) for interval of standard Special
Relativity. In general case integral of motion $\sigma$ does not
equal to $+1$ or $-1$. Specifically, an account of internal
structure of free mass point gives

$$
\eta_{\mu\nu} dR^{\mu} dR^{\nu} + \frac{1}{m_0}
\varepsilon_{\mu\nu\lambda\kappa} S^{\lambda\kappa} dR^{\mu}
dU^{\nu} = \left[ 1 + \frac{1}{m_0 c}
\frac{\varepsilon_{\lambda\kappa\tau\omega} S^{\lambda\kappa}
U^{\tau} W^{\omega}}{\eta_{\rho\sigma} U^{\rho} U^{\sigma}}
\right] \eta_{\mu\nu} dR^{\mu} dR^{\nu} = \sigma d\lambda^2 \;,
\eqno{({\rm III}.30)}
$$

\noindent i.e. the Minkowski space-time ${\bf E}_{1,3}^{\rm R}$
effectively extends to 8-dimensional phase space with interval
(III.30) and degenerate metric, which is equivalent to
4-dimensional conformally flat space with metric

$$
g_{\mu\nu} = \left[ 1 + \frac{1}{m_0 c}
\frac{\varepsilon_{\lambda\kappa\tau\omega} S^{\lambda\kappa}
U^{\tau} W^{\omega}}{\eta_{\rho\sigma} U^{\rho} U^{\sigma}}
\right] \eta_{\mu\nu} \;, \eqno{({\rm III}.31)}
$$

\noindent coordinate dependence of which may be determined, as
soon as solution of equation of motion (III.24) for $U = 0$ is
found.

\end{document}